\documentclass[aps,prx,twocolumn,nofootinbib]{revtex4-2}
\usepackage{amsmath}
\usepackage{amsfonts}
\usepackage{graphicx}
\usepackage{color}
\usepackage{dsfont}
\usepackage{verbatim}
\usepackage{mathtools}
\usepackage[normalem]{ulem}
\usepackage{comment}
\usepackage{stackrel}
\usepackage{booktabs}
\usepackage{bm}
\usepackage[pdftex]{hyperref}
\hypersetup{colorlinks = true, urlcolor = black, linkcolor = black, citecolor = black}

\definecolor{myblue}{rgb}{.93, .93, 1}

\newcommand{\bsub}{\begin{subequations}}
	\newcommand{\esub}{\end{subequations}}

\begin{document}
	
	\title{Correlated insulator in two Coulomb-coupled quantum wires}
	
	\author{Yang-Zhi~Chou}\email{yzchou@umd.edu}
	\affiliation{Condensed Matter Theory Center and Joint Quantum Institute, Department of Physics, University of Maryland, College Park, Maryland 20742, USA}
	
	\author{Sankar Das~Sarma}
	\affiliation{Condensed Matter Theory Center and Joint Quantum Institute, Department of Physics, University of Maryland, College Park, Maryland 20742, USA}	
	
	\date{\today}
	
	\begin{abstract}
		Motivated by the recently discovered incompressible insulating phase in the bilayer graphene exciton experiment [arXiv:2306.16995], we study using bosonization two Coulomb-coupled spinless quantum wires and examine the possibility of realizing the similar phenomenology in one dimension. We explore the possible phases as functions of $k_{F}$'s and interactions. We show that an incompressible insulating phase can arise for two lightly doped electron-hole quantum wires  (i.e., $k_{F1}=-k_{F2}$ and small $|k_{F1}|$) due to strong interwire interactions. Such an insulating phase forms a parity-even wire-antisymmetric charge density wave without interwire phase coherence, which melts to a phase allowing for a perfect negative drag upon heating. The finite-temperature response is qualitatively consistent with the ``exciton solid'' phenomenology in the bilayer graphene exciton experiment.
	\end{abstract}
	
	\maketitle
	
	\section{Introduction}
	
	Low-dimensional quantum many-body systems have been an active research area because of the unconventional phases and the high tunability in the experimental setups. An exciting direction is to create quantum phases with inter-subsystem correlation and coherence, such as Coulomb drag \cite{Narozhny2016,Yamamoto2006,Laroche2011positive,Laroche2014,Du2021coulomb} and electron-hole bilayer phenomena \cite{Eisenstein2014exciton,Nandi2012,Li2017excitonic,Liu2017quantum,Liu2022crossover,Gu2022dipolar,Zeng2023exciton,Wang2023excitonic,Zeng2023evidence,Davis2023josephson,Bai2022evidence,Nguyen2023perfect,Qi2023perfect}. For example, two layers with opposite charge carriers (one with electrons, one with holes) can form exciton condensates due to the interlayer Coulomb interaction, building up strong interlayer phase coherence and allowing for intriguing observable consequences. A recent bilayer graphene exciton experiment \cite{Zeng2023evidence} observes a correlated insulating state, a putative ``exciton solid.'' Upon heating, such a state melts into an exciton condensate phase, characterized by a perfect negative drag response. It is speculated that this insulating state is distinct from a bilayer Wigner crystal that arises independently in each layer. However, the nature of the putative exciton solid is not well understood.

	Motivated by the correlated insulating state discovered in Ref.~\cite{Zeng2023evidence}, we study the interacting two-subsystem problem in a corresponding one-dimensional (1D) biwire analog, where controlled analytical tools are available. In this work, we consider two isolated clean spinless quantum wires that are coupled through Coulomb interaction. The goal is to examine possible phases akin to the formation of interwire excitons, and the complete single-wire quantum phase diagram \cite{Giamarchi_Book} is not emphasized in this work. With bosonization, the problem is mapped to two decoupled sine-Gordon models (corresponding to wire-symmetric and wire-antisymmetric sectors), and the resulting phase diagram can be obtained analytically. In particular, we find a correlated insulating state when the two wires are in the dilute carrier limit (i.e., very small $k_F$). This insulating state corresponds to a wire-antisymmetric charge density wave (CDW) with even parity, yielding zero counterflow conductivity. With increasing temperature, the antisymmetric gap becomes inactive, allowing for a perfect negative drag and a finite counterflow conductivity. This finite-temperature biwire phenomenology is consistent with the results of the recent bilayer graphene exciton experiment \cite{Zeng2023evidence}. Notably, the insulating phase in the Coulomb-coupled wires has no interlayer phase coherence, i.e., excitons are absent here. It is an interwire correlated incoherent incompressible insulator phase, and it is not an excitonic supersolid. The prediction of the interwire incoherent insulating state is not specific to 1D systems and should apply to higher dimensions.
	Thus, our results provide a potential explanation for the correlated insulating state discovered in the recent bilayer graphene exciton experiment \cite{Zeng2023evidence}.

\begin{figure}[t]
	\includegraphics[width=0.425\textwidth]{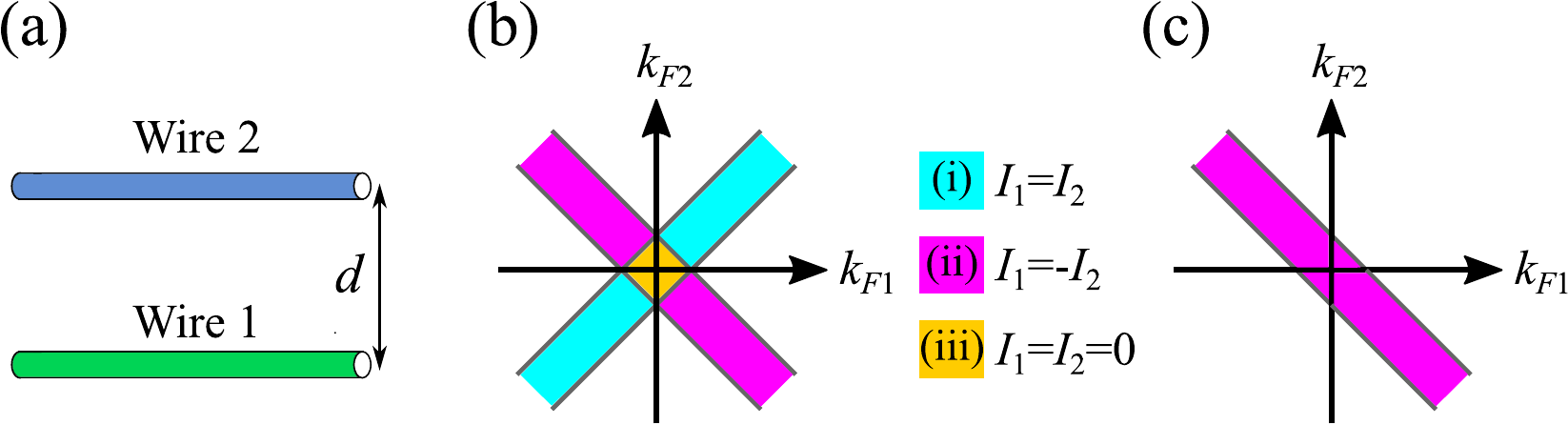}
	\caption{Setup and current correlation diagrams. (a) Two quantum wires are spatially separated with a distance $d$ (i.e., ignoring electron tunneling). Two wires interact through Coulomb interaction. (b) Current correlation diagram of two Coulomb-coupled quantum wires as functions of $k_{F1}$ and $k_{F2}$. We focus only on $K_{\pm}<1$. There are four distinct regions: (i) The cyan regions indicate a positive drag ($I_1=I_2$), (ii) the magenta regions indicate a negative drag ($I_1=-I_2$), (iii) the yellow region indicates an interaction-driven insulator ($I_1=I_2=0$), and (iv) the white regions indicate two decoupled Luttinger liquids, in which $I_1$ and $I_2$ are completely independent. The widths of the regions (i) and (ii) are determined by the $\delta Q_c$ discussed in the main text. (c) Current correlation diagram with $K_{+}<1$, $K_->1$, and infinitesimal $|V_-|$. The wire-antisymmetric sector is always gapless in this case.}
	\label{Fig:Setup_PD}
\end{figure}
	
	\section{Model}
	
	We are interested in two isolated clean spinless quantum wires that are coupled through Coulomb interaction but do not allow any interwire electron tunneling. The setup is related to 1D Coulomb drag  \cite{Nazarov1998,Klesse2000,Fiete2006,Zyuzin10,Chou2015,Kainaris2017,Narozhny2016}, fermion ladders (in the absence of interchain tunneling) \cite{Orignac1997,Giamarchi_Book}, and 1D excitonic insulators \cite{Muljarov2000,Werman2015,Kantian2017,Hu2020} except that we consider general doping densities with arbitrary $k_F$. After linearizing the bands, the physical fermion ($\psi_n$) of the $n$th wire can be decomposed into chiral right ($R_n$) and left mover ($L_n$), described by $\psi_n(x)\approx e^{ik_{Fn}x}R_n(x)+e^{-ik_{Fn}x}L_n(x)$, where $k_{Fn}$ is the Fermi wavevector of the $n$th wire. Positive (negative) $k_{Fn}$ corresponds to electron (hole) doping \cite{CarrierConvention}. We aim to construct a quantum phase diagram with tuning parameters $k_{F1}$, $k_{F2}$, and interaction strength.

	The Coulomb coupled wires can be described by $\hat{H}=\hat{H}_0+\hat{H}_{\text{LL}}+\hat{H}_{I}$, where
	\begin{subequations}\label{Eq:H}
		\begin{align}
			\hat{H}_0=&\sum_{n=1,2}\!v_F\!\int dx\left[R^{\dagger}_n\left(-i\partial_xR_n\right)\!-\!L^{\dagger}_n\left(-i\partial_xL_n\right)\right],\\
			\nonumber\hat{H}_{I}=&V_+\int dx\left[e^{i2Q_+x}L^{\dagger}_1R_1L^{\dagger}_2R_2+\text{H.c.}\right]\\
			&+V_-\int dx\left[e^{i2Q_-x}L^{\dagger}_1R_1R^{\dagger}_2L_2+\text{H.c.}\right],
		\end{align}
	\end{subequations}
	and $\hat{H}_{\text{LL}}$ encodes the Luttinger liquid interactions (both intrawire and interwire processes). In Eq.~(\ref{Eq:H}), $v_F$ is the Fermi velocity, $V_+>0$ and $V_->0$ correspond to the interwire Coulomb-induced backscattering interaction, and $Q_{\pm}\equiv k_{F1}\pm k_{F2}$. The model obeys the spinless time-reversal symmetry and parity symmetry. The time-reversal operation $\mathcal{T}$ corresponds to $R_n\rightarrow L_n$, $L_n\rightarrow R_n$, and $i\rightarrow-i$; the parity operation $\mathcal{P}$ corresponds to $R_n\rightarrow L_n$, $L_n\rightarrow R_n$, and $x\rightarrow-x$. For simplicity, we have assumed that two wires have the same Fermi velocity and the same intrawire interaction, but the values of $k_{F1}$ and $k_{F2}$ are not constrained. We also ignore the intrawire umklapp backscattering interactions as it is typically preempted by the interwire backscattering interactions. Our qualitative results do not rely on these nonessential simplifications.

		\begin{table}[t]
		\centering
		\begin{tabular}{c||c|c|c|c}
			\toprule
			 & (i) & (iia) & (iib) & (iii) \\
			\midrule
				$\psi^{\dagger}_1\psi_2+$H.c.	& $\mathsf{x}$ & $\square$ & \checkmark & $\mathsf{x}$ \\
				$\psi_1\psi_2+$H.c.				& $\square$  & $\mathsf{x}$ & $\mathsf{x}$ & $\mathsf{x}$ \\
				\midrule
				$\mathcal{O}_{\text{CDW},+}$	&$\mathsf{x}$  & $\mathsf{x}$ & $\mathsf{x}$ & $\mathsf{x}$ \\
				$\mathcal{O}_{\text{CDW},-}$	& \checkmark & \checkmark  & $\square$ & \checkmark \\
				\midrule
				$\mathcal{D}_{\text{CDW},+}$	& $\mathsf{x}$ & \checkmark  & $\square$ & $\mathsf{x}$ \\
				$\mathcal{D}_{\text{CDW},-}$	& \checkmark & $\mathsf{x}$ & $\mathsf{x}$ & $\mathsf{x}$ \\
			\bottomrule
		\end{tabular}
		\caption{Order parameters in various cases based on the power-law exponents. \checkmark indicates the dominant order parameter based on the nonlocal correlation function. $\square$ indicates the subleading order parameter. $\mathsf{x}$ indicates vanishingly small expectation value. For region (ii), we separate case (iia) ($K_-<1$ and small $|Q_-|$) and case (iib) ($K_->1$ or large $|Q_-|$) as the results are different. For region (iii), $\mathcal{O}_{\text{CDW},-}$ is a constant, suggesting a true long-range order due to discrete symmetry breaking of the minima in $V_{+}$ and $V_-$ terms in Eq.~(\ref{Eq:H_b_pm}).}
		\label{tab:Instability}
	\end{table}

	Now, we discuss the interaction effects. The Luttinger liquid interactions renormalize $\hat{H}_0$ and turn the system into (gapless) Luttinger liquids, allowing for bosonic excitations \cite{Giamarchi_Book}. The interwire backscattering interactions can induce correlation gaps. To examine the interaction effects systematically, we employ bosonization \cite{Shankar_Book,Giamarchi_Book}. The interacting fermionic Hamiltonian $\hat{H}$ is then mapped to a bosonic Hamiltonian, given by
	\begin{align}\label{Eq:H_b}
		\nonumber\hat{H}_{b}=&\sum_{n=1,2}\int dx\frac{v}{2\pi}\left[K\left(\partial_x\phi_n\right)^2+\frac{1}{K}\left(\partial_x\theta_n\right)^2\right]\\
		\nonumber&-\frac{V_+}{2\pi^2\alpha^2}\int dx\cos\left[2(\theta_1+\theta_2)+2Q_+x\right]\\
		\nonumber&+\frac{V_-}{2\pi^2\alpha^2}\int dx\cos\left[2(\theta_1-\theta_2)+2Q_-x\right]\\
		&+\frac{V'}{\pi^2}\int dx\left(\partial_x\theta_1\right)\left(\partial_x\theta_2\right)
	\end{align} 
	where $\phi_n$ and $\theta_n$ are the phase and density bosons respectively, $v$ is the velocity, $K$ is the Luttinger parameter ($K<1$ for repulsive interaction), $V'>0$ is the strength of the interwire Luttinger liquid interaction. Note that the Luttinger parameter $K=1/\sqrt{1+\frac{U'}{\pi v_F}}$, where $U'$ encodes the intrawire Luttinger liquid interaction strength. Again, we have assumed the same Fermi velocity and intrawire interaction among the two wires. In this convention, the long-wavelength density operator and the current operator are expressed by $\rho_n=\frac{1}{\pi}\partial_x\theta_n$ and $I_n=-\frac{1}{\pi}\partial_t\theta_n$, respectively. Note that the minus sign in the $V_+$ term is due to the bosonization convention used in this work. See Appendix~\ref{App:bosonization} for a discussion.
	
	The bosonic Hamiltonian [Eq.~(\ref{Eq:H_b})] can be simplified with a change of basis. We introduce collective bosonic fields, $\Phi_{\pm}=\frac{1}{\sqrt{2}}\left[\phi_1\pm\phi_2\right]$ and $\Theta_{\pm}=\frac{1}{\sqrt{2}}\left[\theta_1\pm\theta_2\right]$. The subscript  $+$ ($-$) indicates the wire-symmetric (wire-antisymmetric) degrees of freedom. With these new variables, equation~(\ref{Eq:H_b}) becomes $\hat{H}_b\rightarrow\hat{H}_{b+}+\hat{H}_{b-}$, where
	\begin{subequations}\label{Eq:H_b_pm}
		\begin{align}
			\nonumber\hat{H}_{b,+}=&\int dx\frac{v_+}{2\pi}\left[K_+\left(\partial_x\Phi_+\right)^2+\frac{1}{K_+}\left(\partial_x\Theta_+\right)^2\right]\\
			&-\frac{V_+}{2\pi^2\alpha^2}\int dx\cos\left[2\sqrt{2}\Theta_++2Q_+x\right],\\
			\nonumber\hat{H}_{b,-}=&\int dx\frac{v_-}{2\pi}\left[K_-\left(\partial_x\Phi_-\right)^2+\frac{1}{K_-}\left(\partial_x\Theta_-\right)^2\right]\\
			&+\frac{V_-}{2\pi^2\alpha^2}\int dx\cos\left[2\sqrt{2}\Theta_-+2Q_-x\right].
		\end{align}
	\end{subequations}
	In the above expression, $v_+$ and $v_-$ are the velocities, and $K_+$ and $K_-$ are the Luttinger parameters. One can show that
	\begin{align}\label{Eq:v_pm_K_pm}
		v_{\pm}=v_F\sqrt{1+\frac{U'\pm V'}{\pi v_F}},\,\,\,\, K_{\pm}=1/\sqrt{1+\frac{U'\pm V'}{\pi v_F}}.
	\end{align}

	Thus, $K_+<K_-$  generally holds because the $V'$ term in Eq.~(\ref{Eq:H_b}) contributes to a repulsive interaction in the symmetric sector and an attractive interaction in the antisymmetric sector \cite{Klesse2000}. If the intrawire Luttinger liquid interaction ($U'$) is stronger than the interwire Luttinger interaction ($V'$), one should expect $K_+<K_-<1$ \cite{Klesse2000,Chou2019Loc_driven}. We also consider the case $K_+<1<K_-$, corresponding to $U'<V'$ \cite{InterwireK}.
	$\hat{H}_{b+}$ and $\hat{H}_{b-}$ are completely decoupled, and each of them corresponds to the sine-Gordon model which is mathematically equivalent to the commensurate-incommensurate transition problem \cite{PokrovskyTalapov,Giamarchi_Book,Chou2019Loc_driven}. In addition, the bosonic Hamiltonian $\hat{H}_{b,\pm}$ can be mapped to noninteracting massive Dirac fermion at the Luther-Emery point \cite{Giamarchi_Book,Chou2019Loc_driven} ($K_{\pm}=1/2$ in this case), providing an intuitive way to analyze the results through gapped fermionic insulators. Note that the qualitative results obtained from Luther-Emery analysis also apply for general cases with $K_{\pm}<1$. See Appendix~\ref{App:LE} for a detailed discussion about Luther-Emery refermionization. Our goal is to construct the quantum phase diagram arising from the interplay between interaction and doping (i.e., $k_{F1}$ and $k_{F2}$).

\begin{figure}[t]
	\includegraphics[width=0.25\textwidth]{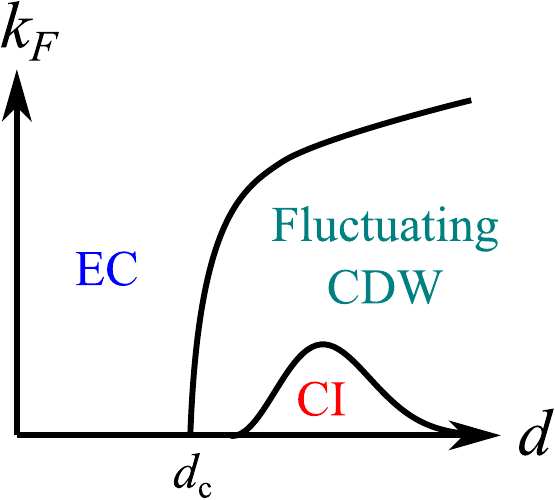}
	\caption{Phase diagram of the electron-hole two-wire system. We sketch the phase diagram as functions of wire separation $d$ and $k_F$. CI indicates the correlated insulator phase with $I_1=I_2=0$; EC denotes the exciton (quasi)condensate phase ($I_1=-I_2$); Fluctuating CDW denotes the negative drag phase ($I_1=-I_2$) with leading (power-law decaying) CDW instability. $d_c$ separates $K_->1$ ($d<d_c$) and $K_-<1$ regimes. See main text for detailed discussion.}
	\label{Fig:eh_phase}
\end{figure}
	
	\section{Possible phases and instabilities in general Coulomb-coupled two-wire problem}
	
	With bosonization and the collective variables, the two Coulomb-coupled quantum wires can be mapped to decoupled $\hat{H}_{b,+}$ and $\hat{H}_{b,-}$, which are related to the sine-Gordon model. In the following, we summarize the main results and construct the quantum phase diagram of the two Coulomb-coupled quantum wires. See Appendix~\ref{App:OPs} for a discussion on technical details.

	The bosonized Hamiltonian $\hat{H}_{b,\pm}$ [Eq.~(\ref{Eq:H_b_pm})] can be mapped to the commensurate-incommensurate transition problem \cite{PokrovskyTalapov}. The $V_{\pm}$ term in $\hat{H}_{b,\pm}$ can be ignored (i.e., a Luttinger liquid phase, incommensurate) if $|2Q_{\pm}|>\delta Q_c$, where $\delta Q_c$ indicates the threshold value for the commensurate-incommensurate transition \cite{PokrovskyTalapov} depending on the interaction parameters. When $|2Q_{\pm}|<\delta Q_c$, $\hat{H}_{b,\pm}$ can be viewed as the commensurate sine-Gordon model (with $Q_{\pm}=0$), which admits the same renormalization group (RG) flow as the Berezinskii–Kosterlitz–Thouless transition \cite{Giamarchi_Book}. For $K_{\pm}<1$, the $V_{\pm}$ term in $\hat{H}_{b,\pm}$ becomes relevant (in the RG sense), and the system develops a gap. For $K_{\pm}>1$, the system can develop a gap only when $|V_{\pm}|$ is larger than the critical value (corresponding to the separatrix of the transition). Otherwise, the system is described by a gapless Luttinger liquid.

	For $K_{\pm}<1$, there are four distinct regions as illustrated in Fig.~\ref{Fig:Setup_PD}(b): (i) A gapless phase with a positive current drag $I_1=I_2$ (the cyan regions, $k_{F1}\approx k_{F2}$), (ii) a gapless phase with a negative current drag $I_1=-I_2$ (the magenta regions, $k_{F1}\approx-k_{F2}$), (iii) a correlation-induced incompressible insulator (the yellow region, $k_{F1}, k_{F2}\approx 0$) \cite{Chou2019Loc_driven}, and (iv) decoupled quantum wires (the white region, generic $k_{F1},k_{F2}$). For $K_+<1$ but $K_->1$, the value of $|V_-|$ is important in determining the phases. Thus, the existence of regions (i) and (iii) depends on whether $|V_-|$ is above the critical threshold. For infinitesimal $|V_-|$, the region (i) is absent, and the region (iii) is replaced by region (ii) with $I_1=-I_2$ as shown in Fig.~\ref{Fig:Setup_PD}(c).

	We introduce several order parameters to examine the possible phases in Fig.~\ref{Fig:Setup_PD}(b). The interwire correlation can be described by the interwire exciton pair (e.g., $\psi^{\dagger}_1\psi_2$) and interwire Cooper pair (e.g., $\psi_1\psi_2$) operators. In addition, we consider charge density wave (CDW) operators, $\mathcal{O}_{\text{CDW},\pm}$ and $\mathcal{D}_{\text{CDW},\pm}$. $\mathcal{O}_{\text{CDW},+}$ and $\mathcal{O}_{\text{CDW},-}$ ($\mathcal{D}_{\text{CDW},+}$ and $\mathcal{D}_{\text{CDW},-}$) correspond to wire-symmetric and wire-antisymmetric CDW. $\mathcal{O}_{\text{CDW},\pm}$ is parity-even, while $\mathcal{D}_{\text{CDW},\pm}$ is parity-odd. All the CDW operators obey spinless time-reversal symmetry. Due to the Mermin-Wagner theorem, the continuous symmetry breaking is absent in infinite 1D systems. We thus examine the nonlocal correlation functions of the order parameters, and the slowest decaying correlation function determines the dominant instability \cite{Giamarchi_Book}. The detailed analysis is discussed in Appendix~\ref{App:OPs}. We summarize the main results in the following (also in Table~\ref{tab:Instability}).
	
	In the region (i) ($I_1=I_2$), the $\mathcal{O}_{\text{CDW},-}$ and $\mathcal{D}_{\text{CDW},-}$ are the leading coexisting order parameters, the system tends to form an interlocked CDW instability \cite{Klesse2000,Starykh2000gapped,Narozhny2016,Furuya2015}. In the region (ii) ($I_1=-I_2$) with $K_{\pm}<1$ and small $|Q_-|$, the leading order parameters are coexisting $\mathcal{O}_{\text{CDW},-}$ and $\mathcal{D}_{\text{CDW},+}$, corresponding to an interlayer electron-hole CDW instability \cite{Furuya2015}. When $K_->1$ or $|Q_-|$ is sufficiently large, the exciton pair operator becomes dominant \cite{EC}, implying a formation of exciton (quasi)condensate. Note that the CDW and exciton operators are not compatible, i.e., they cannot acquire finite expectation values simultaneously. Thus, one should expect a phase transition within the region (ii). Note that the instabilities discussed in regions (i) and (ii) exhibit power-law correlations, implying the absence of true long-range orders.
	In region (iii), the order parameter $\mathcal{O}_{\text{CDW},-}$ is the only nontrivial order parameter and is constant (corresponding to a discrete symmetry breaking due to $V_+$ and $V_-$ interaction). The ground state is described by a wire-antisymmetric parity-even CDW order. The results discussed above are summarized in Table~\ref{tab:Instability}.

	\begin{figure}[t]
	\includegraphics[width=0.45\textwidth]{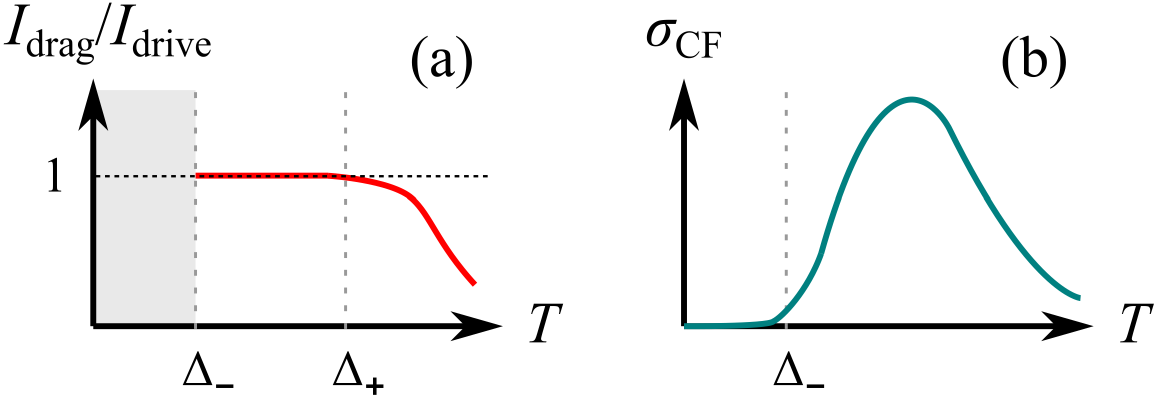}
	\caption{Finite-temperature drag and counterflow conductivity of correlated insulators. (a) The sketched drag ratio, $I_{\text{drag}}/I_\text{drive}$. $I_\text{drive}$ denotes the current in the active wire with a current injected; $I_{\text{drag}}$ denotes the current in the passive wire and is due to Coulomb interaction from the active wire. This quantity is ill-defined for $T<\Delta_-$ as $I_1$ and $I_2$ are essentially zero. (b) The sketched counterflow conductivity ($\sigma_{\text{CF}}$). See main text for a detailed discussion.}
	\label{Fig:Response}
\end{figure}

	\section{Phase diagram of electron-hole bi-wire system}
	
	The main goal of this work is to investigate the possible connection to the exciton systems. Thus, we consider $k_{F1}=-k_{F2}=k_F>0$ (corresponding to an electron-hole bi-wire system). A schematic phase diagram is sketched in Fig.~\ref{Fig:eh_phase}, where we incorporate the long-range Coulomb potential for the interwire interactions. In the following, we discuss the qualitative results in Fig.~\ref{Fig:eh_phase}. Technical details can be found in Appendix.~\ref{App:phase_diagram}.
	
	First, the Luttinger parameters $K_+$ and $K_-$ depend on the interwire interaction, which is controlled by $d$. The interwire interaction $V'(d)\approx\frac{2e^2}{\kappa}K_0(2\pi d/L)$, where $\kappa$ is the dielectric constant, $K_0$ is the zeroth order modified Bessel function of the second kind, $L$ is the wire length. With Eq.~(\ref{Eq:v_pm_K_pm}) and the expression of $V'$, the Luttinger parameter in the antisymmetric sector is given by
		\begin{align}
			K_-=\left[1+\frac{U'}{\pi v_F}-\frac{2e^2}{\pi v_F \kappa}K_0(2\pi d/L)\right]^{-1/2}.
		\end{align}
	One can show that $K_+$ ($K_-$) decreases as $d$ decreases (increases). We define a critical value $d_c$ such that $U'=V'(d_c)$ corresponding to $K_-=1$. In the bi-wire systems with $k_{F1}=-k_{F2}=k_F>0$, the symmetric sector is always gapped as $K_+<1$. Thus, the phase diagram is primarily controlled by $K_-$ (which depends on $d$) and $k_F$.
	
	For an infinitesimal $k_F$, the exciton condensate phase is realized for $K_->1$. For $K_-<1$, the exciton condensate can become the leading instability if $k_F$ is sufficiently large. As shown in Eq.~(\ref{S_Eq_OP_ii}), the exciton pair correlation is non-oscillating (carrying zero wavevector), while the CDW correlation functions oscillate with a wavevector $|Q_-|=2k_F$. The rapid oscillation in CDW correlation suppresses the susceptibility, which is the Fourier transform of the space-time correlation function. Therefore, the exciton condensate phase is favored for large $k_F$. The correlated insulator phase is realized when both symmetric and antisymmetric sectors are gapped and can be viewed as a stable electron-hole CDW state. This correlated insulator phase only exists for a sufficiently small $k_{F}$ (doping) and intermediate $d$. The commensuration condition requires that $4v_-k_F<\Delta_-^{(0)}$, where $\Delta_-^{(0)}$ is the antisymmetric correlation gap with $k_F=0$. The nonmonotonic behavior in $d$ can be understood through the $d$-dependence of $K_-$ and $V_-$. Using the Gaussian variational method \cite{Giamarchi_Book} for $K_-<1$, the correlation gap (with $k_F=0$) is given by
		\begin{align}
			\Delta_-^{(0)}= v_-\Lambda\left(\frac{4K_-V_-}{\pi v_-\Lambda^2\alpha^2}\right)^{\frac{1}{2-2K_-}},
		\end{align}
	where $\alpha$ is the ultraviolet length scale and $\Lambda$ is the energy cutoff.
	To achieve $K_-<1$, $d$ must be larger than $d_c$. However, $V_-\approx \frac{2e^2}{\kappa}K_0(4k_Fd)$ decays as $d$ increases. 
	As a result, the correlated insulating phase is the strongest for an intermediate $d$. See Appendix~\ref{App:phase_diagram} for an extended discussion.

	\section{Finite-temperature phenomenology of correlated insulating phase}
	
	In this section, we focus on the correlated insulating phase and study the finite-temperature response (negative drag ratio and counterflow conductivity). There are two distinct energy scales $\Delta_+$ and $\Delta_-$ corresponding to the correlation gap in the symmetric and antisymmetric sectors respectively, and $\Delta_-<\Delta_+$ generally.

	First, we inspect the negative drag. To understand the temperature dependence, we discuss the response in three asymptotic regimes: $T\ll\Delta_-$, $\Delta_-\ll T\ll \Delta_+$, and $T\gg \Delta_+$. At low temperatures ($T\ll\Delta_-$), currents in both wires vanish, and the drag ratio is not well defined. A perfect negative current drag (i.e., $I_\text{drive}=I_{\text{drag}}$) is developed when $\Delta_-\ll T\ll \Delta_+$. As the temperature increases, we expect that $\Delta_+$ correlation weakens, and the drag ratio decreases monotonically in the high temperature limit. In reality, $\Delta_+$ might not be much larger than $\Delta_-$, but we can still infer the finite-temperature behavior by extrapolating the three asymptotic limits. 
	
	The counterflow conductivity is a direct probe of the antisymmetric sector as the applied voltage is opposite between the two subsystems. Thus, the counterflow conductivity here is mathematically similar to the conductivity of a bosonized sine-Gordon model, corresponding to a fermionic gapped insulator with a chemical potential inside the gap. 
	For $T\ll \Delta_-$, we expect an exponentially small activated conductivity, which can be obtained by performing the dilute-instanton calculations \cite{Rajaraman1982solitons} or through mapping to Luther-Emery fermions (see Appendix~\ref{App:LE}). 
	We also expect a high-temperature suppression of conductivity due to inelastic scatterings \cite{Chou2015} (e.g., intrawire umklapp interaction, which is an irrelevant perturbation and ignored in our zero-temperature analysis). Then, we interpolate the two asymptotic limits and construct the nonmonotonic finite-temperature conductivity as shown in Fig.~\ref{Fig:Response}(b).
	
	With the results mentioned above, we sketch the finite-temperature negative drag ratio and counterflow conductivity in Fig.~\ref{Fig:Response}. Remarkably, the results are qualitatively similar to the experimentally observed correlated insulated state \cite{Zeng2023evidence}.

	\section{Discussion}
	
	We study two Coulomb-coupled quantum wires and analyze the possible quantum phases. Strikingly, we show that the finite-temperature behavior of the incoherent correlated insulating interwire CDW is qualitatively consistent with the phenomenology of the putative ``exciton solid'' phase in the bilayer graphene exciton experiment \cite{Zeng2023evidence}. Based on our theory, the insulating state is a stable electron-hole CDW state, mimicking an interwire excitonic state. However, the interwire phase coherence (defined by the exciton pair correlation) is absent in this interwire CDW state. The correlated insulating state here is not specific to 1D systems, and we expect interlayer Coulomb interactions can generate similar correlated insulating states in higher-dimensional electron-hole bilayer systems. Our results suggest a possible non-exciton interpretation of the correlated insulating phase observed in the experiment \cite{Zeng2023evidence}.
	
	Other incompressible insulating ground states have been proposed in literature \cite{Demler2001,Yang2001,Vu2023excitonic}. In particular, closely bound excitons can form an interlayer electron-hole Wigner crystal state \cite{Yang2001}, which is adiabatic to two independently formed intralayer Wigner crystals being locked by interlayer interaction. Note that our predicted electron-hole CDW state becomes two decoupled Luttinger liquids when the interwire interactions are completely suppressed, indicating a very different mechanism from the Wigner crystal picture. Another difference is that disorder-pinning is not required for realizing insulating states in our theory as the effective band structure is fully gapped. These distinct features suggest that the correlated insulating state emphasized in this work is qualitatively different from the electron-hole Wigner crystal \cite{Yang2001}.

	In this work, the existence of the correlated insulating state is due to the gapping out of symmetric and antisymmetric sectors of the bi-wire problem based on Luttinger liquid analysis. It is possible to generalize our work to a higher dimension. For example, one can build the two-dimensional electron and hole layers by coupling an array of 1D quantum wires \cite{Mukhopadhyay2001} and then consider the Coulomb interaction between these two layers. When each layer realizes a sliding Luttinger liquid \cite{Mukhopadhyay2001}, the analysis of our work can straightforwardly apply, and the correlated insulating state can be realized with a similar condition. Note that the two-dimensional coupled-wire model is highly anisotropic and stems from 1D systems. Constructing an isotropic two-dimensional model that allows for the correlated insulating state is an interesting future direction.
	
	One potential issue about the correlated insulating state is the requirement of small $k_F$. Particularly, the standard Luttinger liquid theory, which is built on linearized dispersion, is not strictly valid in the presence of strong nonlinear dispersion corrections (e.g., near the bottom of a $k^2$ dispersing band). We note that this is a quantitative issue. First, the commensuration condition requires $k_F<k_c=\Delta_-^{(0)}/(4v_-)$, and enhancing interwire interaction can increase the value of $k_c$. Note that $k_c$ depends on energy cutoff $\Lambda$ which also decreases as $k_F$ decreases in a $k^2$ dispersing band. In addition, there exist 1D systems with approximately linear-in-$k$ dispersion, such as topological edge states. The predicted correlated insulating state is more likely to be found in quantum wires with linear-in-$k$ dispersion and strong interwire interaction.

	Now, we discuss the stability of the results in the presence of disorder. If the disorder potential is smooth and merely induces long-wavelength fluctuation, the situation is mathematically equivalent to two disordered helical edge states studied in Ref.~\cite{Chou2019Loc_driven}. The main difference is that the critical value of Luttinger parameter is reduced to $3/4$ from $1$, and the fully gapped states are replaced by localized insulating states. For generic disorder (including backscattering processes), we expect Anderson localization for long wires (wire length $L$ longer than the localization length $l_{\text{loc}}$). For short wires with strong Coulomb interactions ($v_F\tau_c\ll L\ll l_{\text{loc}}$, where $\tau_c$ is the correlation time due to interwire Coulomb interaction), the predictions of this work should apply.

	\begin{acknowledgments}
		We thank Jay D. Sau for useful discussion. This work is supported by the Laboratory for Physical Sciences.
	\end{acknowledgments}

	\appendix
	
		\section{Microscopic Model}\label{App:Model}
	
	In this appendix, we introduce a concrete microscopic model that corresponds to the long-wavelength model discussed in Eq.~(\ref{Eq:H}).
	We consider two spatially separated quantum wires that are interacting through the Coulomb force. The Hamiltonian can be described as follows:
	\begin{align}
		\nonumber\hat{H}_{\text{2-wire}}=&\sum_{n=1,2}\sum_{k}\left[\epsilon_{n}(k)-\mu_n\right]\psi^{\dagger}_{n}(k)\psi_{n}(k)\\
		\nonumber&+\frac{1}{2}\sum_{n=1,2}\int dxdx'\mathcal{U}_n(x-x')\rho_n(x)\rho_n(x')\\
		&+\int dx dx'\mathcal{V}(x-x')\rho_1(x)\rho_2(x'),
	\end{align}
	where $n$ is the wire index, $\psi_n$ denotes the annihilation operator for fermion, $\epsilon_{n}(k)$ is the dispersion, $\mu_n$ is the chemical potential, $\mathcal{U}_n$ is the intrawire interaction potential, $\mathcal{V}$ is the interwire Coulomb potential, $\rho_n(x)=\psi_n^{\dagger}(x)\psi_n(x)$ is the density operator. The intrawire interaction contains long-range and short-range component of the Coulomb interaction, while the interwire interaction is dictated by the long-range Coulomb interaction. 
	
	In the low-energy limit, we approximate the bands by linear dispersions with constant Fermi velocity. In addition, we decompose the fermion field by $\psi_n\approx e^{ik_{Fn}}R_n+e^{-ik_{Fn}}L_n$. The Hamiltonian becomes $\hat{H}=\hat{H}_0+\hat{H}_{\text{int}}$, where
	\begin{align}
		\hat{H}_0=&\sum_{n=1,2}v_{Fn}\int dx\left[R^{\dagger}_n\left(-i\partial_xR_n\right)-L^{\dagger}_n\left(-i\partial_xL_n\right)\right],\\
		\nonumber\hat{H}_{\text{int}}\approx&\sum_{n=1,2}\left\{\hat{H}^{(\text{intra})}_{\text{LL},n}\!+\!U_n\!\int dx\left[\!e^{i4k_{Fn}x}\!:\!\left(L^{\dagger}_nR_n\right)^2\!:\!+\text{H.c.}\right]\!\right\}\\
		\nonumber&+V'\int dx\left[R^{\dagger}_1R_1+L^{\dagger}_1L_1\right]\left[R^{\dagger}_2R_2+L^{\dagger}_2L_2\right]\\
		\nonumber&+V_+\int dx\left[e^{i2Q_+x}L^{\dagger}_1R_1L^{\dagger}_2R_2+\text{H.c.}\right]\\
		&+V_-\int dx\left[e^{i2Q_-x}L^{\dagger}_1R_1R^{\dagger}_2L_2+\text{H.c.}\right].
	\end{align}
	In the above expression, $Q_{\pm}=k_{F1}\pm k_{F2}$, $\hat{H}^{(\text{intra})}_{\text{LL},n}$ denotes the intrawire Luttinger liquid interaction of the $n$th wire, $V'$ terms is the interwire Luttinger liquid interaction, $U_n$ and $V_{\pm}$ are the intrawire and interwire backscattering interactions.
	We consider phenomenological Luttinger liquid (short-range) interactions instead of the long-range Coulomb interaction in $\hat{H}_{\text{intra LL},n}$. This is a standard approximation. Inclusion of the long-range Coulomb interaction does not alter our qualitative results. In the main text, we ignore the intrawire backscattering $U_n$ term because it is generally preempted by the interwire backscattering terms. We also assume $v_{F1}=v_{F2}$ for simplicity.

	\section{Bosonization convention}\label{App:bosonization}

	Now, we introduce the bosonization convention used in this work. The chiral fermions can be expressed by
	\begin{align}
		R_n=\frac{\eta_n}{\sqrt{2\pi\alpha}}e^{i(\phi_n+\theta_n)},\,\,\,L_n=\frac{\eta_n}{\sqrt{2\pi\alpha}}e^{i(\phi_n-\theta_n)},
	\end{align}
	where $\eta_n$ is the Klein factor for the $n$th wire, $\phi_n$ is the phase boson, $\theta_n$ is the density boson, and $\alpha$ is the ultraviolet length scale in our theory. The bosons obey the following commutation relation $\left[\phi_n(x),\theta_{n'}(y)\right]=-i\delta_{n,n'}\pi u(y-x)$,
	where $u(z)$ is the Heaviside function. This particular choice enables anticommutation relation of fermions within the same wire. The long-wavelength component of the density in the $n$th wire is expressed by $\rho_n=\frac{1}{\pi}\partial_x\theta_n$.
	
	Fermion bilinears can be computed using the following identities,
	\begin{align}
		\label{M1}e^{A}e^{B}&=:e^{A+B}:e^{\frac{1}{2}\langle A^2+B^2+2AB\rangle},\\
		\label{M2}e^{A}e^{B}&=e^{A+B}e^{\frac{1}{2}[A,B]}=e^Be^Ae^{[A,B]}.
	\end{align}
	Particularly,
	\begin{align}
		R^{\dagger}_n\left(-i\partial_xR_n\right)\rightarrow&\frac{1}{4\pi}\left(\partial_x\phi_n+\partial_x\theta_n\right)^2,\\
		-L^{\dagger}_n\left(-i\partial_xL_n\right)\rightarrow&\frac{1}{4\pi}\left(\partial_x\phi_n-\partial_x\theta_n\right)^2,\\
		L^{\dagger}_nR_n\rightarrow&\frac{-i}{2\pi\alpha}e^{i2\theta_n},\\
		R^{\dagger}_nL_n\rightarrow&\frac{i}{2\pi\alpha}e^{-i2\theta_n}
	\end{align}

	With bosonization, the two Coulomb-coupled quantum wires can be expressed by
	\begin{align}
		\nonumber\hat{H}_{b}=&\sum_{n=1,2}\int dx\frac{v_n}{2\pi}\left[K_n\left(\partial_x\phi_n\right)^2+\frac{1}{K_n}\left(\partial_x\theta_n\right)^2\right]\\
		\nonumber&-\sum_{n=1,2}\frac{U_n}{2\pi^2\alpha^2}\int dx\cos\left[4\theta_n+4k_{Fn}x\right]\\
		\nonumber&-\frac{V_+}{2\pi^2\alpha^2}\int dx\cos\left[2(\theta_1+\theta_2)+(2k_{F1}+2k_{F2})x\right]\\
		\nonumber&+\frac{V_-}{2\pi^2\alpha^2}\int dx\cos\left[2(\theta_1-\theta_2)+(2k_{F1}-2k_{F2})x\right]\\
		\label{S:Eq:H_b}&+\frac{V'}{\pi^2}\int dx \left(\partial_x\theta_1\right)\left(\partial_x\theta_2\right).
	\end{align}
	Note that the minus sign in front of $U_n$ and $V_+$ terms are due to the bosonization convention used here. The velocity ($v_n$) and Luttinger parameter ($K_n$) are given by
	\begin{align}
		v_n=v_{Fn}\sqrt{1+\frac{U'_n}{\pi v_F}},\,\,\,K_n=1/\sqrt{1+\frac{U'_n}{\pi v_F}},
	\end{align}
	where $U'_n$ is the phenomenological interaction parameter from $\hat{H}_{\text{intra LL},n}$. 
	Since $U'_n>0$, we obtain $K_n<1$.
	
	For simplicity, we assume $v_1=v_2=v$ and $K_1=K_2=K$. With these approximations, the model can be further simplified by introducing the collective variables as follows:
	\begin{align}
		\Phi_{\pm}=\frac{1}{\sqrt{2}}\left[\phi_1\pm\phi_2\right],\,\,\,\Theta_{\pm}=\frac{1}{\sqrt{2}}\left[\theta_1\pm\theta_2\right],
	\end{align}
	where the subscript $+$ means the wire-symmetric modes and the subscript $-$ means the wire-antisymmetric modes.
	
	We ignore the $U_n$ term as it is generally preempted by $V_+$ and.or $V_-$. The bosonic Hamiltonian can be expressed by $\hat{H}_b=\hat{H}_{b,+}+\hat{H}_{b,-}$, where
	\begin{subequations}\label{S_Eq:H_pm}
		\begin{align}
			\nonumber\hat{H}_{b,+}=&\int dx\frac{v_+}{2\pi}\left[K_+\left(\partial_x\Phi_+\right)^2+\frac{1}{K_+}\left(\partial_x\Theta_+\right)^2\right]\\
			&-\frac{V_+}{2\pi^2\alpha^2}\int dx\cos\left[2\sqrt{2}\Theta_++2Q_+x\right],\\
			\nonumber\hat{H}_{b,-}=&\int dx\frac{v_-}{2\pi}\left[K_-\left(\partial_x\Phi_-\right)^2+\frac{1}{K_-}\left(\partial_x\Theta_-\right)^2\right]\\
			&+\frac{V_-}{2\pi^2\alpha^2}\int dx\cos\left[2\sqrt{2}\Theta_-+2Q_-x\right].
		\end{align}
	\end{subequations}
	In the above expression, the $V'$ term in Eq.~(\ref{S:Eq:H_b}) has been incorporated. Using $(\partial_x\theta_1)(\partial_x\theta_2)=\frac{1}{2}\left[\left(\partial_x\Theta_+\right)^2-\left(\partial_x\Theta_-\right)^2\right]$, we obtain
	\begin{align}
		v_{\pm}=v_F\sqrt{1+\frac{U'\pm V'}{\pi v_F}},\,\,\,\, K_{\pm}=1/\sqrt{1+\frac{U'\pm V'}{\pi v_F}}.
	\end{align}
	Thus, $K_+<K_-$ holds as long as $V'>0$. $K_-$ can become larger than 1 if $V'>U'$.
	The results here suggest a complete factorization of the symmetric and antisymmetric degrees of freedom. Thus, the two Coulomb-coupled wires can be viewed as two decoupled sine-Gordon models.

	\begin{figure}[t]
		\includegraphics[width=0.35\textwidth]{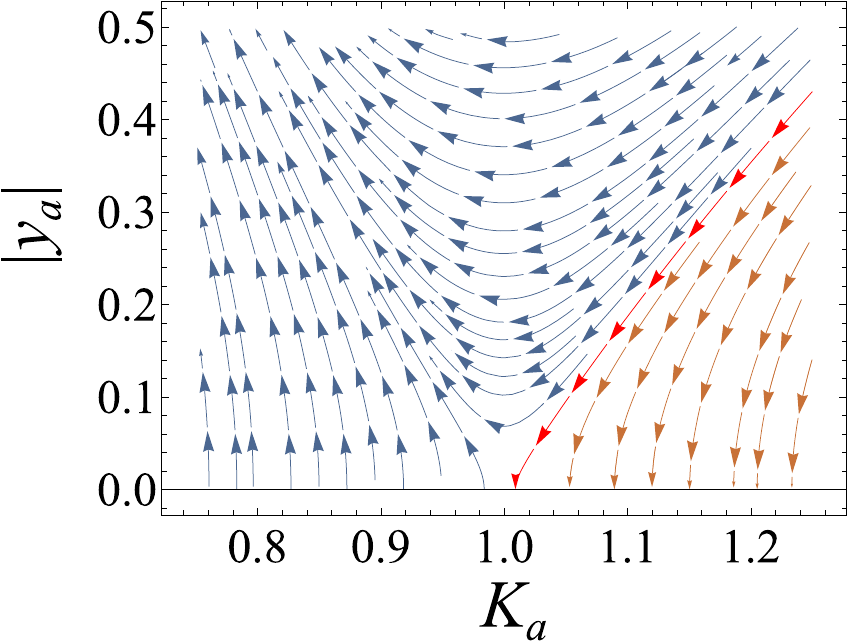}
		\caption{Renormalization group flows for sine-Gordon model. The subscript $a=+,-$ indicates the symmetric and antisymmetric sectors. The red arrows indicate the separatrix which divides two distinct regimes. The brown arrows indicate flows to the fixed line of Luttinger liquid with $K_a\ge 1$. The blue arrows indicate flows to strong coupling limit in which $|y_a|$ diverges.}
		\label{Fig:RG}
	\end{figure}
	
	\section{Refermionization at Luther-Emery point}\label{App:LE}
	
	At $K_{\pm}=1/2$, equation~(\ref{Eq:H_b_pm}) can be mapped to massive Dirac fermions. To see this, we first introduce $\tilde{\Phi}_{\pm}/\sqrt{2}=\Phi_{\pm}$ and $\tilde{\Theta}_{\pm}=\sqrt{2}\Theta_{\pm}+Q_{\pm}x$. Equation~(\ref{Eq:H_b_pm}) with $K_{\pm}=1/2$ becomes
	\begin{align}
		\nonumber\hat{H}_{b,+}=&\int dx\frac{v_+}{2\pi}\left[\left(\partial_x\tilde\Phi_+\right)^2+\left(\partial_x\tilde\Theta_+\right)^2\right]\\
		&-\frac{v_+Q_+}{\pi}\int dx (\partial_x\tilde\Theta_+)-\frac{V_+}{2\pi^2\alpha^2}\int dx\cos\left[2\tilde\Theta_+\right],\\
		\nonumber\hat{H}_{b,-}=&\int dx\frac{v_-}{2\pi}\left[\left(\partial_x\tilde\Phi_-\right)^2+\left(\partial_x\tilde\Theta_-\right)^2\right]\\
		&-\frac{v_-Q_-}{\pi}\int dx (\partial_x\tilde\Theta_-)+\frac{V_-}{2\pi^2\alpha^2}\int dx\cos\left[2\tilde\Theta_-\right].
	\end{align}
	Now, we introduce Luther-Emery fermions given by
	\begin{align}
		\Phi_{R,\pm}=\frac{1}{\sqrt{2\pi\alpha}}e^{i(\tilde{\Phi}_{\pm}+\tilde\Theta_{\pm})},\,\,\,\Phi_{L,\pm}=\frac{1}{\sqrt{2\pi\alpha}}e^{i(\tilde{\Phi}_{\pm}-\tilde\Theta_{\pm})}.
	\end{align}
	Note that Klein factors are not needed here as symmetric and antisymmetric sectors are complete decoupled in our case. Finally, we map the bosonic Hamiltonian to the massive Dirac fermions.
	\begin{align}
		\nonumber\hat{H}_{b,\pm}\rightarrow &v_\pm\int dx\left[\Psi^{\dagger}_{R,\pm}\left(-i\partial_x\Psi_{R,\pm}\right)-\Psi^{\dagger}_{L,\pm}\left(-i\partial_x\Psi_{L,\pm}\right)\right]\\
		\nonumber&-\mu_+\int dx\left[\Psi^{\dagger}_{R,\pm}\Psi_{R,\pm}+\Psi^{\dagger}_{L,\pm}\Psi_{L,\pm}\right]\\
		\label{Eq:App:H_LM}&\mp M_\pm\int dx\left[i\Psi^{\dagger}_{L,\pm}\Psi_{R,\pm}-i\Psi^{\dagger}_{R,\pm}\Psi_{L,\pm}\right],
	\end{align}
	where $\mu_{\pm}=v_{\pm}Q_\pm$ and $M_\pm=V_\pm/(2\pi\alpha)$. While the Luther-Emery fermions are not simply related to the physical fermions, the density and current operators are related upto a factor of 2. Thus, we can use this noninteracting fermion theory to compute conductivity of the original problem.
	
	Using the fermionic Hamiltonian in Eq.~(\ref{Eq:App:H_LM}), we can infer several results in the main text. We focus on $|\mu_\pm|<M_\pm$, where a zero-temperature state is gapped with semiconductor-like dispersion. For $0<T\ll M_{\pm}$, several physical quantities (such as conductivity) can be understood by the thermal activated exponential behavior. For $T>M_{\pm}$, the conductivity increases as temperature increases. However, the conductivity should not grow without bound. At sufficiently high temperatures, interactions (e.g., intrawire umklapp interaction) that are not included in the Hamiltonian can contribute to inelastic scattering and suppress conductivity. Thus, we expect a monotonic decreasing finite-temperature conductivity in the high-temperature regime. Combining the low and high temperature limits, we can construct a nonmonotonic finite-temperature conductivity as shown in Fig.~\ref{Fig:Response}(b).

	\section{Renormalization group flows}\label{App:RG}

	Here we briefly review the famous renormalization group (RG) flows of $\hat{H}_{b,+}$ and $\hat{H}_{b,-}$ given by Eq.~(\ref{S_Eq:H_pm}). The $Q_{\pm}x$ factor in the cosine term is crucial for the $V_{\pm}$ interaction. This is related to the commensurate-incommensurate transition studied by Pokrovsky and Talapov \cite{PokrovskyTalapov}. There exists a critical threshold $\delta Q_c$ separating the commensurate ($|Q_{\pm}|<\delta Q_c$) and incommensurate ($|Q_{\pm}|>\delta Q_c$) phases. When the system is commensurate, the qualitative result is similar to $Q_{\pm}=0$. In the incommensurate phase, $V_{\pm}$ is irrelevant, and the low-energy theory is described by gapless Luttinger liquid. 
	
	Now, we review the RG flows for the $Q_{\pm}=0$ case. The RG flows is the same as the Berezinskii–Kosterlitz–Thouless transition given by \cite{Giamarchi_Book} 
	\begin{align}
		\frac{d K_{\pm}}{dl}=&-\frac{y_{\pm}^2K^2_{\pm}}{2},\\
		\frac{d y_{\pm}}{dl}=&(2-2K_{\pm})y_{\pm},
	\end{align}
	where $y_{\pm}=V_{\pm}/(\pi v_{\pm})$. In Fig.~\ref{Fig:RG}, we plot the above RG flow equations. For $K_{\pm}<1$, $dy_{\pm}/dl$ is always positive, implying a relevant RG flow for $V_{\pm}$. For $K_{\pm}>1$, $V_{\pm}$ is irrelevant for an infinitesimal $|y_{\pm}|$, but $V_{\pm}$ can become relevant for a $|y_{\pm}|$ exceeding some threshold value ($\approx 2(K_{\pm}-1)$ for $K-1\rightarrow 0^+$, marked by the red arrows in Fig.~\ref{Fig:RG}). When $V_{\pm}$ becomes relevant, the low-energy theory is described by a gapped phase. Otherwise, one should expect a gapless Luttinger liquid phase.

	\section{Order parameters}\label{App:OPs}
	
	In this section, we introduce the order parameters and compute the non-local correlation. The goal is to find the most singular (i.e., less decaying) correlation corresponding to the dominant instability of the system. We consider interwire exciton pair ($\psi^{\dagger}_1\psi_2+\text{H.c.}$), interwire Cooper ($\psi_1\psi_2+\text{H.c.}$), wire-symmetric CDW ($\mathcal{O}_{\text{CDW},+}$ and $\mathcal{D}_{\text{CDW},+}$), and wire-antisymmetric CDW ($\mathcal{O}_{\text{CDW},-}$ and $\mathcal{D}_{\text{CDW},-}$). $\mathcal{O}_{\text{CDW},+}$ and $\mathcal{O}_{\text{CDW},-}$ are parity-even, while $\mathcal{D}_{\text{CDW},+}$ and $\mathcal{D}_{\text{CDW},-}$ are parity-odd. The order parameters are expressed by
	\begin{widetext}
			\begin{align}
			\psi^{\dagger}_1\psi_2+\text{H.c.}\approx&e^{i(-k_{F1}+k_{F2})x}R^{\dagger}_1R_2+e^{i(-k_{F1}-k_{F2})x}R^{\dagger}_1L_2+e^{i(k_{F1}+k_{F2})x}L^{\dagger}_1R_2+e^{i(k_{F1}-k_{F2})x}L^{\dagger}_1L_2+\text{H.c.},\\
			\psi_1\psi_2+\text{H.c.}\approx&e^{i(k_{F1}+k_{F2})x}R_1R_2+e^{i(k_{F1}-k_{F2})x}R_1L_2+e^{i(-k_{F1}+k_{F2})x}L_1R_2+e^{i(-k_{F1}-k_{F2})x}L_1L_2+\text{H.c.},\\
			\mathcal{O}_{\text{CDW},+}=&e^{i2k_{F1}x}L^{\dagger}_1R_1+e^{-i2k_{F1}x}R^{\dagger}_1L_1+ e^{i2k_{F2}x}L^{\dagger}_2R_2+ e^{-i2k_{F2}x}R^{\dagger}_2L_2,\\
			\mathcal{O}_{\text{CDW},-}=&e^{i2k_{F1}x}L^{\dagger}_1R_1+e^{-i2k_{F1}x}R^{\dagger}_1L_1- e^{i2k_{F2}x}L^{\dagger}_2R_2- e^{-i2k_{F2}x}R^{\dagger}_2L_2,\\
			\mathcal{D}_{\text{CDW},+}=&ie^{i2k_{F1}x}L^{\dagger}_1R_1-ie^{-i2k_{F1}x}R^{\dagger}_1L_1+ ie^{i2k_{F2}x}L^{\dagger}_2R_2 - ie^{-i2k_{F2}x}R^{\dagger}_2L_2,\\
			\mathcal{D}_{\text{CDW},-}=&ie^{i2k_{F1}x}L^{\dagger}_1R_1-ie^{-i2k_{F1}x}R^{\dagger}_1L_1- ie^{i2k_{F2}x}L^{\dagger}_2R_2 + ie^{-i2k_{F2}x}R^{\dagger}_2L_2.
		\end{align}
	
	Using bosoniztion, the order parameters become
	\begin{align}
		\psi^{\dagger}_1\psi_2+\text{H.c.}\rightarrow&\frac{\eta_1\eta_2}{\pi\alpha}e^{-i\sqrt{2}\Phi_-}\left[\cos\left(\sqrt{2}\Theta_-+Q_-x\right)+\cos\left(\sqrt{2}\Theta_++Q_+x\right)\right]+\text{H.c.},\\
		\psi_1\psi_2+\text{H.c.}\rightarrow&\frac{\eta_1\eta_2}{\pi\alpha}e^{i\sqrt{2}\Phi_+}\left[\cos\left(\sqrt{2}\Theta_-+Q_-x\right)+\cos\left(\sqrt{2}\Theta_++Q_+x\right)\right]+\text{H.c.},\\
		\mathcal{O}_{\text{CDW},+}\rightarrow&\frac{2}{\pi\alpha}\sin\left(\sqrt{2}\Theta_++Q_+x\right)\cos\left(\sqrt{2}\Theta_-+Q_-x\right),\\
		\mathcal{O}_{\text{CDW},-}\rightarrow&\frac{2}{\pi\alpha}\cos\left(\sqrt{2}\Theta_++Q_+x\right)\sin\left(\sqrt{2}\Theta_-+Q_-x\right),\\
		\mathcal{D}_{\text{CDW},+}\rightarrow&\frac{2}{\pi\alpha}\cos\left(\sqrt{2}\Theta_++Q_+x\right)\cos\left(\sqrt{2}\Theta_-+Q_-x\right),\\
		\mathcal{D}_{\text{CDW},-}\rightarrow&-\frac{2}{\pi\alpha}\sin\left(\sqrt{2}\Theta_++Q_+x\right)\sin\left(\sqrt{2}\Theta_-+Q_-x\right),
	\end{align}
	
	Now, we are in the position to examine the correlation function. We consider four different cases, corresponding to the regions (i), (ii), and (iii) discussed in the main text. 
		\end{widetext}
	
	\subsection{Region (i)}
	
	For region (i) ($I_1=I_2$), the symmetric sector is gapless, and the antisymmetric sector is gapped. The ground state configuration must minimize the $V_-$, i.e., $\cos(2\sqrt{2}\Theta_-)=-1$. Thus, we obtain $\Theta_-=\frac{2m+1}{2\sqrt{2}}\pi$, where $m$ is an integer. Note that a constant value of $\Theta_-$ implies that any operator involving $\Phi_-$ is strongly disordered. The value of $\Theta_+$ also eliminate some of the order parameters. We summarize the nontrivial power-law correlation as follows:
	\begin{subequations}\label{S_Eq_OP_i}
		\begin{align}
			\langle\psi_2^{\dagger}(x,t)\psi_1^{\dagger}(x,t)\psi_1(0,t)\psi_2(0,t)\rangle\propto &\frac{\cos\left(Q_+x\right)}{|x|^{K_++K_+^{-1}}},\\
			\langle \mathcal{O}_{\text{CDW},-}(x,t)\mathcal{O}_{\text{CDW},-}(0,t)\rangle\propto &\frac{\cos\left(Q_+x\right)}{|x|^{K_+}},\\
			\langle \mathcal{D}_{\text{CDW},-}(x,t)\mathcal{D}_{\text{CDW},-}(0,t)\rangle\propto &\frac{\cos\left(Q_+x\right)}{|x|^{K_+}}.
		\end{align}
	\end{subequations}
	The correlation functions of other order parameters are decaying faster than power law, so we ignore. We conclude that $\mathcal{O}_{\text{CDW,-}}$ and $\mathcal{D}_{\text{CDW,-}}$ are the leading instabilities.

	\subsection{Region (ii)}
	
	For region (ii) ($I_1=-I_2$), the symmetric sector is gapped, and the antisymmetric sector is gapless.  The ground state configuration must minimize the $V_+$ term, i.e., $\cos(2\sqrt{2}\Theta_+)=1$. (Note the minus sign in the expression of the $V_+$ term.) Thus, we obtain $\Theta_+=\frac{m}{\sqrt{2}}\pi$, where $m$ is an integer. Since $\Theta_+$ acquires a finite expectation value, the operators involving $\Phi_+$ are strongly disordered. We obtain the leading contributions as follows:
	\begin{subequations}\label{S_Eq_OP_ii}
		\begin{align}
			\langle\psi_2^{\dagger}(x,t)\psi_1(x,t)\psi_1^{\dagger}(0,t)\psi_2(0,t)\rangle\propto &\frac{1}{|x|^{K_-^{-1}}},\\
			\langle \mathcal{O}_{\text{CDW},-}(x,t)\mathcal{O}_{\text{CDW},-}(0,t)\rangle\propto &\frac{\cos\left(Q_-x\right)}{|x|^{K_-}},\\
			\langle \mathcal{D}_{\text{CDW},+}(x,t)\mathcal{D}_{\text{CDW},-}(0,t)\rangle\propto &\frac{\cos\left(Q_-x\right)}{|x|^{K_-}}.
		\end{align}
	\end{subequations}
	The correlation functions of other order parameters are decaying faster than power law, so we ignore. For $K_-<1$, $\mathcal{O}_{\text{CDW,-}}$ and $\mathcal{D}_{\text{CDW,-}}$ are the leading instabilities. For $K_->1$, the interwire exciton pair, $\psi_1^{\dagger}\psi_2$, is the leading instability.

	\subsection{Region (iii)}
	
	For region (iii) ($I_1=I_2=0$), both symmetric and antisymmetric sectors are gapped. The ground state configuration must simultaneously minimize $V_+$ and $V_-$ terms, i.e., $\cos(2\sqrt{2}\Theta_+)=1$ and $\cos(2\sqrt{2}\Theta_-)=-1$. Therefore, we obtain $\Theta_+=\frac{m}{\sqrt{2}}\pi$ and $\Theta_-=\frac{2m'+1}{2\sqrt{2}}\pi$, where $m$ and $m'$ are integers. Most of the order parameters we considered are strongly disordered. The only exception is $\mathcal{O}_{\text{CDW},-}$, which is a constant in this region. Note that this corresponds to a discrete spontaneous symmetry breaking, and a true long-range order is formed.

	\section{Phase diagram as functions of $d$ and $k_F$ for two electron-hole wires}\label{App:phase_diagram}
	
	In this work, we use $U'$ and $V'$ to represent the intrawire and interwire Luttinger liquid interactions. Microscopically, the interwire interaction $V'$ corresponds to $\tilde{\mathcal{V}}(q\rightarrow 2\pi/L)$, where $L$ is the length of wire. We will keep the $d$-dependence in the following discussion as we are interested in the phase diagram controlled by the wire separation $d$. The intrawire interaction $U'$ contains both the regularized $1/r$ Coulomb interaction and short-range interaction, but we are not interested in the explicit functional dependence. For a model with only long-range Coulomb interaction, we should expect $U'>V'$. In this work, we consider a more general situation including $U'<V'$. Since $K_+<1$ holds regardless of the value of $V'$, we focus only on the value $K_-$, which is now expressed by
	\begin{align}\label{S_Eq:K_-}
		K_-=\left[1+\frac{U'}{\pi v_F}-\frac{2e^2}{\pi v_F \kappa}K_0(2\pi d/L)\right]^{-1/2},
	\end{align}
	where $\kappa$ is the dielectric constant and $K_0$ is the zeroth order modified Bessel function of the second kind.
	
	We focus on two electron-hole Coulomb-Coupled wires, corresponding to $k_{F1}=-k_{F2}=k_F>0$. According to Eq.~(\ref{S_Eq:K_-}), a larger $d$ corresponds to a smaller $K_-$. If the system contains significant short-range intrawire attraction (e.g., due to phonons), $K_-$ can become less than 1 for a sufficiently small $d$. Thus, we can define a critical value $d_c$ such that $K_-<1$ for $d>d_c$ and $K_->1$ for $d<d_c$. Note that the $K_->1$ regime might not be accessible if $U'$ is sufficiently large (strong intrawire repulsion). 
	
	Now, we discuss the phase boundary between the exciton condensate and the fluctuating CDW phases. As we can see in Eq.~(\ref{S_Eq_OP_ii}), the exciton pair correlation is not oscillating, while the CDW correlation functions oscillate with a wavevector $Q_-=2k_F$. The susceptibility corresponds to integration of the space-time correlation function, and a rapid oscillation (i.e., large $k_F$) in the correlation function suppresses the corresponding susceptibility. Thus, much of the $K_-<1$ region is taken over by the exciton condensate phase as long as $k_F$ is sufficiently large.
	
	The existence of correlated insulating state depends on whether the antisymmetric sector is gapped. In the two electron-hole wires, the antisymmetric sector corresponds to a commensurate-incommensurate problem. If $K_-<1$ and $2Q_-=4k_F$ is smaller than the critical threshold $\delta Q$, the antisymmetric sector becomes gapped. Otherwise, the antisymmetric sector remains gapless. The value of $\delta Q$ is associated with the size of gapped for a commensurate cosine term. Thus, we can estimate the gap $\Delta_-^{(0)}$ (with $k_F=0$) using the Gaussian variational approach \cite{Giamarchi_Book}:
	\begin{align}\label{S_Eq:Delta_-}
		\Delta_-^{(0)}= v_-\Lambda\left(\frac{4K_-y_-}{\Lambda^2\alpha^2}\right)^{\frac{1}{2-2K_-}},
	\end{align}
	where $\Lambda$ is the energy cutoff.
	In the above expression, the quantity inside the parentheses is less than one. Thus, for a fixed value of $V_-$, a smaller $K_-<1$ results in a larger $\Delta_-^{(0)}$. In our case, both $V_-$ and $K_-$ depend on $d$. A larger $d$ results in a smaller $K_-$ (enhancing $\Delta_-^{(0)}$) and a small $V_-$ (suppressing $\Delta_-^{(0)}$). As a result, we find that the $\Delta_-$ is maximized at an intermediate value of $d$. Based on the results discussed above, we sketch the possible phase diagrams as functions of $k_F$ and $d$.
	
	For $k_F=0$, it is possible to develop a finite $\Delta_-$ for $K_->1$ provided that $V_-$ is sufficiently large. The threshold value of $y_-=V_-/(\pi v_-)$ is approximately $2(K_--1)$ (estimated in the vicinity of $K_-=1$). We find that this can be satisfied quite generally as long as $U'\ge 0$. Thus, we conclude that correlated insulating state can exist for $d<d_c$ and very small $k_F$, which is a very narrow regime in the phase diagram. In fact, $k_F=0$ case realizes a correlated insulator as long as $U'>0$, but the value of $\Delta_-$ may be exponentially small for a very large $d$ (small $V_-$) or a very small $d$ (large $K_-$). In the main text, we focus on small but nonzero $k_F$. The complication of $k_F$ is most likely not relevant to any experimental situation. Therefore, we do not emphasize this subtle situation in main text.
	
	\section{Coupled-wire model}
	
	We discuss a coupled-wire model that allows for a correlated insulating state due to inter-system Coulomb interaction. This model should provide ideas about generalizing our bi-wire system in main text to higher dimension.
	

\begin{thebibliography}{43}
\bibitem[{Car()}]{CarrierConvention}
		\bibinfo{note}{Note that an electron carrier with $-k_F$ is equivalent to a
			hole carrier with $k_F$. In this work, we use electron carriers for both
			wires. The choice of carrier does not affect any of the results.}
\bibitem[{EC()}]{EC}
		\bibinfo{note}{Exciton pair operator can become the leading instability if
			$k_F$ is sufficiently large even with $K_-<1$. Exciton pair correlation is
			non-oscillating, while the CDW correlation functions oscillate with a
			wavevector $|Q_-|$. The rapid oscillation suppresses the susceptibility which
			is the Fourier transform of the space-time correlation function.}
\end{thebibliography}

\end{document}